\documentclass[a4paper,12pt]{article}
\usepackage{a4wide}
\usepackage{verbatim}
\usepackage{bbm}
\usepackage{pstricks}
\usepackage{amsmath}
\usepackage{amssymb}
\usepackage{epsfig}
\usepackage{graphicx}
\usepackage{titlesec}
\usepackage[numbers,sort&compress]{natbib}
\usepackage{color}
\usepackage[
  font={footnotesize} 
  ,bf             
  ,hang           
  ]{caption}
\usepackage[
  colorlinks=true
  ,urlcolor=blue
  ,anchorcolor=black
  ,citecolor=black
  ,filecolor=black
  ,linkcolor=black
  ,menucolor=black
  ,pagecolor=black
  ,linktocpage=true
  ,pdfproducer=medialab
  ,pdfa=true
  ]{hyperref}

\titleformat*{\section}{\large\bfseries}
\titleformat*{\subsection}{\normalsize\bfseries}
\titleformat*{\subsubsection}{\normalsize\bfseries}

\newcommand\email[1]{%
  \begingroup
  \renewcommand\thefootnote{}\footnote{\href{mailto:#1}{#1}}%
  \addtocounter{footnote}{-1}%
  \endgroup
}

\numberwithin{equation}{section}

\author{\bf\normalsize Kristian Holsheimer}
\date{}

\begin{document}
\parindent=0pt
\parskip=12pt
\itemindent=4pt

\begin{center}\thispagestyle{empty}
{\ }\vspace{1cm}

{\bf\Large On the Marginally Relevant Operator\\ in $z=2$ Lifshitz Holography}

\vspace{2.5cm}

{\bf\normalsize Kristian Holsheimer}
\email{k.holsheimer@uva.nl}

{\it\small Institute of Physics, University of Amsterdam\\
Science Park 904, Postbus 94485, 1090 GL Amsterdam, The Netherlands}
\end{center}

\vspace{1.5cm}

\begin{abstract}\vspace{-12pt}\noindent
We study holographic renormalization and RG flow in a strongly-coupled Lifshitz-type theory in 2+1 dimensions with dynamical exponent $z=2$. The bottom-up gravity dual we use is 3+1 dimensional Einstein gravity coupled to a massive vector field. This model contains a marginally relevant operator around the Lifshitz fixed point. We show how holographic renormalization works in the presence of this marginally relevant operator without the need to introduce explicitly cutoff-dependent counterterms. A simple closed-form expression is found for the renormalized on-shell action. We also discuss how asymptotically Lifshitz geometries flow to AdS in the interior due to the marginally relevant operator. We study the behavior of the renormalized entanglement entropy and confirm that it decreases monotonically along the Lifshitz-to-AdS RG flow.
\end{abstract}

\newpage
\tableofcontents



\newpage
\section{Introduction}

Over the past decade, there has been quite some interest in extending gauge/gravity dualities to theories that are not conformally invariant but still exhibit some kind of scaling symmetry. The scaling symmetry that we consider in this note is so-called Lifshitz-type scaling, where time scales differently from space, $x\to\lambda x$ while $t\to\lambda^zt$. The parameter $z$ is the so-called dynamical critical exponent. This anisotropic scaling symmetry is realized geometrically in holography and the resulting geometry is called Lifshitz spacetime \cite{Kachru:2008yh,Koroteev:2007yp}. Holography for these spacetimes does not yet stand on the same firm footing as ordinary AdS spacetimes, and it is therefore interesting to figure out to what extent the usual AdS/CFT techniques apply to these Lifshitz spacetimes.

As our gravity dual we take 3+1 dimensional Einstein gravity coupled to a massive vector \cite{Taylor:2008tg}, also known as the Einstein--Proca theory. In this note we are particularly interested in the case where the dynamical exponent is equal to the number of spatial boundary dimensions, i.e.\ $z=d_s=2$. In this case one finds that the solutions of the Einstein--Proca field equations contain a logarithmic branch. In particular, the leading behavior of the metric is no longer simply a power of the radial coordinate; it contains leading logarithms. In \cite{Cheng:2009df,Park:2012mn} it was argued that these leading logs are related to having a marginally relevant operator in the system. We will build on this observation, although our method of renormalizing the on-shell action will be radically different. In particular, we show that one can renormalize the on-shell action by adding only local counterterms without the need to introduce explicit dependence on the radial cutoff.

In order to perform holographic renormalization, one needs to specify boundary conditions for the fields. In the Hamilton--Jacobi (HJ) formalism, there is a natural way to fix these boundary conditions by fixing the radial scaling of the fields. This input is restrictive enough to find the renormalized on-shell action, while at the same time it is lenient enough to allow for the leading logarithms. For simplicity, we assume translational invariance in the boundary directions. Our method for renormalizing the on-shell action is based on the results of \cite{Baggio:2011cp}, though the special case of $z=2$ will bring some interesting new features. In particular, we find that the renormalized on-shell action will be a non-analytic function of the only Lorentz scalar one can construct at the non-derivative level: the square of the massive vector.

Before we discuss holographic renormalization we first construct a holographic RG flow that interpolates between a Lifshitz-like fixed point in the UV and a conformally invariant fixed point (AdS$_4$). This Lifshitz-to-AdS flow can be seen as a result of turning on the marginally relevant operator. We check that it makes sense to view AdS as an IR solution by studying the renormalized entanglement entropy, where the entangling surface is a strip. This was proposed as a nice candidate $c$-function in \cite{Myers:2012ed} for $d\geq3$ boundary dimensions. Namely, we will check whether the renormalized entanglement entropy decreases monotonically as one follows the RG flow from the Lifshitz-like fixed point to the conformally invariant fixed point.

This paper is organized as follows. In the next section we give a brief overview of some interesting properties of the massive vector model. We explain why the case where the dynamical exponent is equal to the number of (boundary) spatial dimensions ($z=d_s$) is special. In Section \ref{sec:entanglement-entropy} we study the Lifshitz-to-AdS flow using the renormalized entanglement entropy of a strip, which we compute holographically. In the process, we derive a nice and simple expression for this holographic renormalized entanglement entropy. Finally, Section \ref{sec:holographic-renormalization} contains the main discussion of this paper: holographic renormalization in the presence of the marginally relevant operator.

\section{The Massive-Vector Model}

In order to holographically describe a quantum theory that exhibits Lifshitz-like scaling, we should have a geometry that has this scaling symmetry as isometries. Lifshitz spacetime \cite{Kachru:2008yh} is such a geometry; it is given by
\begin{align}\label{eq:lifshitz-metric}
ds^2\ =\ -\text{e}^{2z\,r}\,dt^2+\text{e}^{2r}\,d\vec{x}{\,}^2+dr^2
\end{align}
A shift in the radial coordinate $r\to r+\ell\log\lambda$ generates an anisotropic scaling transformation $(t,x)\to(\lambda^zt,\lambda x)$. For future reference, we mention that the physical scale $\mu$ of the dual field theory is related to the radial coordinate as $\mu\sim \text{e}^r$, so when we talk about power-law divergences we mean $\sim\text{e}^{\#r}$, while terms like $r^\#$ we call logarithmic. As our gravity dual we take Einstein gravity coupled to a massive vector. For other models based on dimensional reduction of the well-established AdS$_5$ holography, see e.g.\ \cite{Chemissany:2012du,Christensen:2013lma,Christensen:2013rfa}. The action for our massive vector theory is:\footnote{In our notation, $\sqrt{g}\equiv\sqrt{|\det(g)|}$.}
\begin{align}\label{eq:massive-vector-lagrangian}
I\ =\ \frac1{16\pi G}\int d^{4}x\sqrt{g}\,\left( R-2\Lambda-\frac14F_{\mu\nu} F^{\mu\nu}-\frac{m^2}{2}\,A_\mu A^\mu \right)+\frac1{8\pi G}\int d^{3}x\sqrt{\gamma}\,K\,.
\end{align}
This action was introduced in the context of Lifshitz holography in \cite{Taylor:2008tg}. The Lifshitz geometry is a solution to the field equations derived from this action provided we also turn on the vector,
\begin{align}\label{eq:lifshitz-vector-background}
A\ =\ \sqrt{\tfrac{2(z-1)}{z}}\ \text{e}^{zr}\,dt\,.
\end{align}
The parameters of the theory are related to the parameters of the geometry as $m^2=2z/L^2$ and $\Lambda=(z^2+z+4)/2L^2$. (We had set the curvature length scale $L=1$ in \eqref{eq:lifshitz-metric}). For future reference, we note that the same action also has an AdS solution:
\begin{align}\label{eq:ads-metric}
ds^2\ &=\ \text{e}^{2r/\ell}\,\ell^2\big(\!-dt^2+d\vec{x}{\,}^2\big)+dr^2\,,
&
A &= 0\,.
\end{align}
We picked our coordinates such that the Lifshitz length scale is set to one, which fixes the AdS scale to be $\ell=\sqrt{3/5}$.

\subsection{The special nature of $z=d_s$}

Before we continue our discussion, let us mention some interesting facts about Lifshitz systems with critical values of the dynamical exponent, $z=d_s$, where $d_s$ is the number of spatial dimensions on the field theory side.\footnote{This paper focuses on the massive vector theory; indications of the special nature of $z=d_s$ in Maxwell--dilaton-type models can be found e.g.\ in \cite{Edalati:2012tc,Edalati:2013tma}.}

\textbf{Fixed point of a duality transformation.}
A first hint at the special nature of $z=d_s$ in the massive-vector model comes from the following argument. Consider the values of the mass and cosmological constant that give rise to a Lifshitz geometry with dynamical exponent $z$ in $d_s+2$ bulk dimensions,
\begin{align}\label{eq:Lambda-m}
m(z,\ell)\ &=\ \frac{\sqrt{d_sz}}{\ell}\,, &
\Lambda(z,\ell)\ &=\ -\frac{z^2+(d_s-1)z+d_s^2}{2\ell^2} \,.
\end{align}
In \cite{Bertoldi:2009vn} it was noticed that there is a dual pair $(z',\ell')$ that gives rise to the same $m$ and $\Lambda$, because the above relation is quadratic. Solving $m(z,\ell)=m(z',\ell')$ together with $\Lambda(z,\ell)=\Lambda(z',\ell')$ yields
\begin{align}
z'&=\frac{d_s^2}{z}\,, &
\ell'&=\frac{d_s\ell}{z}\,.
\end{align}
The critical value $z=d_s$ is the unique fixed point for the above duality transformation.

\textbf{A logarithmic branch.}
If we focus on $d_s=2$ for the moment, we know from the perturbative analysis \cite{Ross:2009ar} that a basis for the independent modes can be chosen as follows:
\begin{align}
1\,,&&
\text{e}^{-(z+2)r}\,,&&
\text{e}^{-\frac12(z+2-\beta)r}\,,&&
\text{e}^{-\frac12(z+2+\beta)r}\,,
\end{align}
where $\beta^2=(z+2)^2+8(z-2)(z-1)$. For $z=2$ we see that $\beta=z+2=4$, which means that two pairs of modes will coincide and so a logarithmic branch will emerge. We will now see what this logarithmic branch looks like when we consider solutions to the equations of motion derived from \eqref{eq:massive-vector-lagrangian}.

\textbf{Seemingly bad logs.}
At present, we do not have a closed-form solution of the equations of motion that exhibits the expected logarithmic behavior, so we study two approximate solutions instead. For one, we can look at linearized perturbations around the Lifshitz background \eqref{eq:lifshitz-metric} and \eqref{eq:lifshitz-vector-background}. The other approximate solution is an asymptotic expansion, where one expands in powers of $\text{e}^{-r}$ (and $r^{-1}$) for large values of the radius $r$.

We start with the linearized solution. The linearized field equations of the massive-vector theory were solved some time ago in \cite{Ross:2009ar} for $d_s=2$. For $z=d_s=2$, it was found that a logarithmic mode emerges that seemed to grow quicker than the background mode. For this reason, it was generally expected to be an irrelevant perturbation of the (pure) Lifshitz solution.\footnote{It was not phrased in this precise way. It was said that the mode proportional to $c$ should be switched off so as to satisfy the asymptotically Lifshitz boundary conditions. The latter depends on what specific boundary conditions one has in mind.} The mode (proportional to $c$) appears in the linearized solution as
\begin{align}
-g_{tt}\ &=\ \text{e}^{4r}(1-2c\,r+\ldots)\,, \\
g_{ij}\ &=\ \text{e}^{2r}(1+c\,r+\ldots)\,, \\
A_{t}\ &=\ \text{e}^{2r}(1-c\,(\tfrac12+r)+\ldots)\,.
\end{align}
This looks pretty bad, because it looks like the asymptotics are destroyed by this mode. One can see, however, that the Lorentz scalar $A^2$ and the volume form constructed from these fields do behave nicely, e.g.
\begin{align}
A^2\ &=\ -1+c+\ldots\,, &
\sqrt{g}\ &=\ \text{e}^{4r}\big(1+\ldots\big)\,.
\end{align}
The ellipses denote other linearized modes that are suppressed by powers of $\text{e}^{-4r}$. The mode proportional to $c$ shifts the background value of $A^2$.

Now let us turn to the asymptotic solution, which was first obtained in  \cite{Cheng:2009df}. The leading behavior of the asymptotic solution is:
\begin{align}
-g_{tt}\ &=\ \frac{\text{e}^{4r}}{r^4}\left( 1+O(r^{-1}) \right)\,,\label{eq:asympt-lif-log-1}\\
g_{xx}\ &=\ r^2\text{e}^{2r}\left( 1+O(r^{-1}) \right)\,,\label{eq:asympt-lif-log-2}\\
A_{t}\ &=\ \frac{\text{e}^{4r}}{r^2}\left( 1+O(r^{-1}) \right)\,.\label{eq:asympt-lif-log-3}
\end{align}
In this case, the logarithmic modes look even worse. However, just as in the linearized solution, one sees that $A^2$ and the volume form do behave nicely. In this case they only receive sub-leading logarithmic corrections:
\begin{align}
A^2\ &=\ -1+\frac2r+O(r^{-2})\,, &
\sqrt{|g|}\ &=\ \text{e}^{4r}\big(1+\frac1r+O(r^{-2})\big)\,.
\end{align}
The shift in $A^2$ can thus be seen as a logarithmic correction, which vanishes when $r\to\infty$. It was argued in \cite{Cheng:2009df} that the logarithms $\sim r$ that appear in the asymptotic solution comprise a marginally \emph{relevant} perturbation of the ``pure'' Lifshitz solution \eqref{eq:lifshitz-metric}. Our results agree with this statement, although our method of renormalizing the on-shell action is inherently different. We discuss this at the end of Section \ref{sec:holographic-renormalization}.

In conclusion, we see that a logarithmic branch opened up when $z=d_s=2$. This logarithmic branch looks problematic if one considers quantities that are \emph{not covariant}. However, everything appears fine again once we consider only covariant quantities (and the volume density). In particular, we checked explicitly that the curvature invariants and the geodesic deviation behave in this same way. In light of this, it seems appropriate to call the configuration \eqref{eq:asympt-lif-log-1}--\eqref{eq:asympt-lif-log-3} asymptotically Lifshitz, even though the metric looks quite different from the pure Lifshitz geometry \eqref{eq:lifshitz-metric}. Since the asymptotics are not changed in this covariant sense, one can expect that the logarithmic branch is related to a marginally \emph{relevant} perturbation of the pure-Lifshitz solution. We will make this more precise in the context of holographic renormalization.

\textbf{Tidal forces in the infra-red.}
It was previously argued that the Lifshitz geometry \eqref{eq:lifshitz-metric} is singular in the infra-red. Even though the curvature invariants are finite everywhere, one finds that the tidal forces that a local observer experiences diverge as $\text{e}^r\to0$ whenever $z\neq1$, cf.\  \cite{Copsey:2010ya,Horowitz:2011gh}. The logarithmic Lifshitz solution is free of such singularities for the obvious reason that the dynamical exponent flows to $z=1$ in the infra-red (cf.\ Figure \ref{fig:z_eff}).\footnote{See Appendix \ref{sec:numerics} for the numerical setup.} This is in contrast to what was expected in \cite{Copsey:2010ya}, where it was argued that a sensible IR geometry that is free of these pathologies was unlikely to exist. The reason why the analysis from \cite{Copsey:2010ya} does not apply to this particular flow is that we allow for the presence of leading logs.
\begin{figure}[h]\begin{center}
\includegraphics{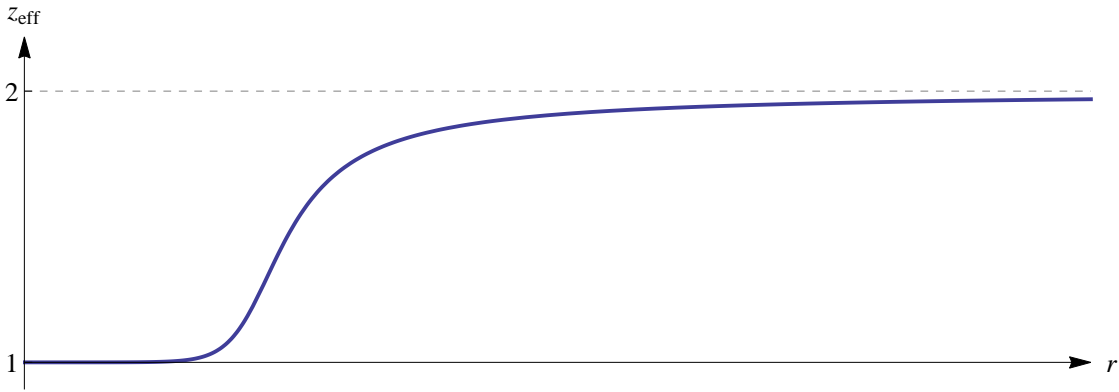}
\caption{The dynamical exponent is evaluated on a numerical background that interpolates between AdS$_4$ in the interior (left) and Lifshitz spacetime in the asymptotic region (right). The dynamical exponent flows from $z_\text{eff}=1$ in the IR to $z_\text{eff}=2$ in the UV.}
\label{fig:z_eff}
\end{center}\end{figure}

\textbf{Weyl anomaly.}
Finally, let us also mention that when $z=d_s$ there is an anomaly akin to the Weyl anomaly in Lorentz-invariant theories. This was first computed in \cite{Adam:2009gq} for $z=d_s=3$. For $z=d_s=2$, the anomaly is given by \cite{Griffin:2011xs,Baggio:2011ha,Chemissany:2012du}:
\begin{align}
\mathcal{A}\ =\ \frac{C_1}{16\pi}\left(K_{ij}K^{ij}-\frac12K^2\right)+\frac{C_2}{16\pi} \left( R-\frac1{N^2}\partial_iN \partial^iN+\frac1{N}\Delta N \right)^2\,,
\end{align}
where $R[h_{ij}]$ and $K_{ij}$ are the intrinsic and extrinsic curvatures on a constant-time slice and $N$ is the associated time lapse. The central charges $C_1$ and $C_2$ depend on the details of the theory; for instance, a free scalar field minimally coupled to the background fields $(N,N_i,h_{ij})$ has $C_1=1/4$ and $C_2=0$. In this note, we only consider systems that are invariant under translations in the boundary direction, i.e.\ $\partial_t=0$ and $\partial_i=0$, so the anomalous breaking of Lifshitz scaling symmetry will play no role at present.

\begin{figure}[h]
\vspace{12pt}
\begin{center}
\scalebox{1} 
{
\begin{pspicture}(0,-1.0089062)(6.09,0.98890626)
\definecolor{color245}{rgb}{0.803921568627451,0.803921568627451,0.803921568627451}
\psbezier[linewidth=0.04,linecolor=color245,fillstyle=solid,fillcolor=color245](5.88,0.88890624)(5.98,0.88890624)(6.06,0.68890625)(5.98,0.38890624)(5.9,0.08890625)(5.94,-0.15109375)(5.94,-0.07109375)(5.94,0.00890625)(5.78,0.88890624)(5.88,0.88890624)
\psbezier[linewidth=0.04,linecolor=color245,fillstyle=solid,fillcolor=color245](0.26,-0.07109375)(0.16,-0.07109375)(0.08,0.12890625)(0.16,0.42890626)(0.24,0.7289063)(0.2,0.9689062)(0.2,0.88890624)(0.2,0.80890626)(0.36,-0.07109375)(0.26,-0.07109375)
\pspolygon[linewidth=0.002,linecolor=color245,fillstyle=solid,fillcolor=color245](0.24,-0.07109375)(0.22,0.9089062)(5.92,0.9089062)(5.92,-0.07109375)
\psline[linewidth=0.036cm]{cc-cc}(0.14,0.92890626)(6.02,0.9089062)
\psline[linewidth=0.04cm]{cc-cc}(0.14,-0.07109375)(6.0,-0.09109375)
\psline[linewidth=0.02cm,arrowsize=0.04cm 2.0,arrowlength=2.0,arrowinset=0.0]{<->}(3.1,0.88890624)(3.1,-0.05109375)
\psline[linewidth=0.02cm,arrowsize=0.04cm 2.0,arrowlength=2.0,arrowinset=0.0]{<->}(6.08,-0.47109374)(0.0,-0.47109374)
\usefont{T1}{ptm}{m}{n}
\rput(3.0814064,-0.7860938){$\widetilde{L}$}
\usefont{T1}{ptm}{m}{n}
\rput(3.4314063,0.39390624){$L$}
\end{pspicture} 
}
\end{center}
\vspace{-20pt}
\caption{The entangling region is a strip. The two length scales associated to this geometry is the width $L$ of the strip and a long-distance cutoff $\widetilde{L}$.}
\label{fig:strip}
\end{figure}
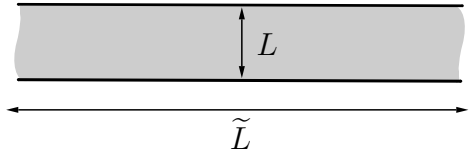
%

\section{Renormalized Entanglement Entropy}\label{sec:entanglement-entropy}

We would like to have a measure for the effective number of degrees of freedom. For this purpose, we shall look at the renormalized entanglement entropy, which was proposed as a candidate $c$-function in \cite{Myers:2012ed} for RG flows that interpolate between conformally invariant fixed points in $d\geq3$ dimensions. We will follow \cite{Myers:2012ed} and study the entanglement entropy associated to a strip-shaped region in flat space (see Figure \ref{fig:strip}). For a strip in $d\ge3$ dimensions, the entanglement entropy contains only the leading area-law divergence and a universal piece \cite{Myers:2012ed,Hsu:2008af}:
\begin{align}\label{eq:entanglement-strip}
S_\text{ent}\ =\ \frac{\text{Area}}{\epsilon^{d-2}}+S_\text{finite}\,,
\end{align}
The strip is particularly convenient, because there are no sub-leading power-law divergences beyond the leading area-law term and $S_\text{finite}$ is independent of the UV-cutoff (so no $\log \epsilon$ dependence).\footnote{For a more generic entangling geometry that is not flat, one finds curvature-dependent power-law divergences that are sub-leading compared to the leading area-law term. Such terms typically do depend on $L$ and even though its dependence can be scaled away by rescaling the UV-cutoff $\epsilon$, it is far simpler to consider a strip for the purpose of finding a quantity that behaves as a $c$-function.} Let $L$ be the width of the strip and let $\widetilde{L}$ be an IR length scale associated to the transverse directions. For conformally invariant fixed points, one finds $S_\text{finite}\propto c_d\,\big(\widetilde{L}/L\big)^{d-2}$, where $c_d$ are the known {\sc a}-type central charges when $d$ is even. The strip geometry is special, because the non-universal power-law divergent piece is independent of the width $L$ of the strip. This means that the universal piece can be extracted rather easily by taking the derivative with respect to the width of the strip $L$,
\begin{align}
S_\text{finite}\ \propto L\,\partial_L S_\text{ent}\,.
\end{align}
The right-hand side we shall call the renormalized entanglement entropy of the strip (in analogy to the renormalized entanglement entropy of the sphere \cite{Liu:2012eea}). In the case where a theory flows between two conformally invariant fixed points, it was suggested in \cite{Myers:2012ed} that the renormalized entanglement entropy would be a good candidate $c$-function:
\begin{align}\label{eq:c-function}
c_d(L)\ =\ \beta_d\,\left(\frac{L}{\widetilde{L}}\right)^{d-2}\,L\,\partial_L S_\text{ent}\,.
\end{align}
The prefactor $\beta_d$ is a dimensionless constant that depends on the number of dimensions as
\begin{align}
\beta_d\ =\ \frac1{\sqrt{\pi}\,2^d\,\Gamma \big( \tfrac{d}{2} \big)}\,\left( \frac{\Gamma \big( \tfrac{1}{2(d-1)} \big)}{\Gamma \big( \tfrac{d}{2(d-1)} \big)} \right)^{d-1}\,.
\end{align}
The function $c_d$ was constructed in such a way that it reduces to the known {\sc a}-type central charges at conformally invariant fixed points. 

In our situation, one of the fixed points we are interested in is not conformally invariant, so it is not a priori clear whether it makes sense to interpret \eqref{eq:c-function} as a $c$-function. However, we will compute \eqref{eq:c-function} holographically, in which case one finds that the computations of the entanglement entropy done either in AdS or in Lifshitz spacetime go through in precisely the same manner. The monotonicity of \eqref{eq:c-function} for non-Lorentz invariant situations was recently discussed in \cite{Cremonini:2013ipa}. One can easily check that our setup meets the requirements of \cite{Cremonini:2013ipa}. In particular, the null energy condition is satisfied, $u_\mu u_\nu T^{\mu\nu}=\frac{m^2}{2}(u^\mu A_\mu)^2\geq0$, where $u^\mu$ is a future-directed null vector. Furthermore, the Ryu--Takayanagi formula holds in the massive vector bulk model, so our computation will be very similar to the known AdS/CFT computations in Einstein gravity.\footnote{One can see that the Wald charge (or improvements thereof) reduces to the area formula in the massive vector model.}

We will use the function \eqref{eq:c-function} to see how the effective number of degrees of freedom decrease along the RG flow. Before we do so, however, we will first derive a simple formula for $c_d(L)$ using holography.

\subsection{A simple holographic formula for the entanglement \textit{c}-function}
Generically it is rather difficult to compute the entanglement entropy away from a scale-invariant fixed point. This why we will use holography. The holographic formula for the entanglement entropy associated to some subregion $A$ was proposed by Ryu and Takayanagi \cite{Ryu:2006bv}. It is given by the area (in Planck units) of a minimal surface in the bulk that is suspended from the boundary $\partial A$ of the subregion $A$.\footnote{This formula is incomplete e.g.\ when higher derivatives are taken into account.} So, the entanglement entropy is given by the on-shell value of the Nambu--Goto type action
\begin{align}
S_\text{ent}\ =\ \frac1{4G_d}\int d^{d-1}x\,\sqrt{\gamma}\,,
\end{align}
where $\gamma_{ab}=g_{\mu\nu}(x)\,\partial_ax^\mu\,\partial_bx^\nu$ is the induced metric on the hypersurface and $G_d$ is Newton's constant in $d$ dimensions. Consider the situation in which the metric at constant time is given by
\begin{align}
ds^2\Big|_{t=\text{const}}\ =\ f(r)\,d\vec{x}{\,}^2+dr^2\,.
\end{align}
Note that this includes both AdS as well as Lifshitz spacetime. Focusing on the case where the entangling subregion is a strip yields
\begin{align}
S_\text{ent}\ &=\ \int_0^L dx\,\mathcal{L}&
\mathcal{L}(r,\dot{r})\ &=\ \frac{\widetilde{L}^{d-2}}{4G_d}\,\sqrt{f(r)\big( f(r)+\dot{r}^2 \big)}
\end{align}
where $\dot{r}=dr/dx$. We have chosen our coordinates such that the strip lies perpendicular to the coordinate $x$ in such a way that it covers the interval $-L<x<L$, so $x$ has been rescaled by a factor of $1/2$ compared to the standard one. Furthermore, we have used the symmetry $x\to-x$, such that the integral runs from $0<x<L$ rather than $-L<x<L$. These two redefinitions generate two factors of $2$, which mutually cancel.
\begin{figure}[!b]
\vspace{6pt}\hrule\vspace{12pt}
\scalebox{1} 
{
\begin{pspicture}(0,-4.324406)(15.11,4.284406)
\definecolor{color1019}{rgb}{0.803921568627451,0.803921568627451,0.803921568627451}
\definecolor{color1027}{rgb}{0.7333333333333333,0.7333333333333333,0.7333333333333333}
\definecolor{color1054}{rgb}{0.5019607843137255,0.5019607843137255,0.5019607843137255}
\psbezier[linewidth=0.04,linecolor=color1019,fillstyle=solid,fillcolor=color1019](8.35,3.6884062)(7.99,3.6884062)(8.73,3.8484063)(8.85,4.008406)(8.97,4.1684065)(9.45,4.2084064)(9.59,4.2084064)(9.73,4.2084064)(8.71,3.6884062)(8.35,3.6884062)
\pspolygon[linewidth=0.002,linecolor=color1019,fillstyle=solid,fillcolor=color1019](8.39,3.6884062)(9.617959,4.2284064)(14.83,4.2284064)(13.85,3.6684062)
\psbezier[linewidth=0.04,linecolor=color1019,fillstyle=solid,fillcolor=color1019](13.85,3.6884062)(14.13,3.6484063)(14.13,3.7284062)(14.35,3.8884063)(14.57,4.048406)(15.09,4.2084064)(14.75,4.2084064)(14.41,4.2084064)(13.57,3.7284062)(13.85,3.6884062)
\psbezier[linewidth=0.03,linecolor=color1027](14.377,4.1884065)(14.25,2.0484064)(13.71,2.2284062)(13.49,3.6484063)
\psdots[dotsize=0.08](13.497,3.6484063)
\psbezier[linewidth=0.03]{cc-cc}(14.35,3.8684063)(14.23,2.3684063)(13.77,2.0284061)(13.49,3.6484063)
\psdots[dotsize=0.072](14.377,4.2084064)
\psline[linewidth=0.02cm]{cc-cc}(9.11,2.6084063)(14.07,2.5684063)
\psline[linewidth=0.036cm]{cc-cc}(9.27,4.2284064)(14.93,4.2084064)
\psline[linewidth=0.04cm]{cc-cc}(8.07,3.6684062)(14.01,3.6484063)
\psbezier[linewidth=0.03,linecolor=color1027](9.737,4.2084064)(9.61,2.0684063)(8.93,2.2684062)(8.71,3.6884062)
\psdots[dotsize=0.08](8.717,3.6684062)
\psbezier[linewidth=0.03]{cc-cc}(9.69,3.6884062)(9.55,2.3284063)(8.99,2.1884062)(8.71,3.6884062)
\psdots[dotsize=0.072](9.737,4.2284064)
\psbezier[linewidth=0.04](7.19,0.78840625)(6.87,-0.39159375)(6.31,-2.4315937)(5.15,-2.4315937)
\psline[linewidth=0.04cm,linecolor=color1054](7.19,0.80840623)(7.25,1.0684062)
\psdots[dotsize=0.08](5.17,-2.4315937)
\psdots[dotsize=0.08](7.19,0.78840625)
\psdots[dotsize=0.08,linecolor=color1054](7.25,1.0684062)
\psbezier[linewidth=0.03,linestyle=dashed,dash=0.16cm 0.16cm](3.07,0.98840624)(3.31,-0.11159375)(3.99,-2.4315937)(5.15,-2.4315937)
\psdots[dotsize=0.06](3.05,1.0884062)
\psline[linewidth=0.02cm,arrowsize=0.05291667cm 3.0,arrowlength=1.4,arrowinset=0.0]{<-}(2.09,1.8684063)(2.09,-4.051594)
\psline[linewidth=0.02cm,arrowsize=0.05291667cm 3.0,arrowlength=1.4,arrowinset=0.0]{->}(1.67,-3.8115938)(9.85,-3.8115938)
\usefont{T1}{ppl}{m}{n}
\rput(2.0545313,2.1984062){$r(x)$}
\usefont{T1}{ppl}{m}{n}
\rput(10.334531,-3.8015938){$x$}
\psbezier[linewidth=0.12,arrowsize=0.05291667cm 4.0,arrowlength=1.4,arrowinset=0.0]{cc->}(8.57,2.8884063)(6.55,3.4484062)(5.77,2.9484062)(5.31,1.1284063)
\psline[linewidth=0.04cm](5.17,-3.8115938)(5.17,-3.8915937)
\psline[linewidth=0.04cm](7.19,-3.8115938)(7.19,-3.8915937)
\psline[linewidth=0.04cm](3.09,-3.8115938)(3.09,-3.8915937)
\psline[linewidth=0.04cm](2.09,0.80840623)(2.01,0.80840623)
\psline[linewidth=0.04cm](2.09,1.1084063)(2.01,1.1084063)
\usefont{T1}{ppl}{m}{n}
\rput(1.6245313,1.2184062){$\infty$}
\usefont{T1}{ppl}{m}{n}
\rput(1.2445313,0.6984063){$-\log\epsilon$}
\usefont{T1}{ppl}{m}{n}
\rput(3.0545313,-4.121594){$-L$}
\usefont{T1}{ppl}{m}{n}
\rput(7.1445312,-4.121594){$L$}
\usefont{T1}{ppl}{m}{n}
\rput(5.134531,-4.121594){$0$}
\psline[linewidth=0.04cm](2.09,-2.4315937)(2.01,-2.4315937)
\usefont{T1}{ppl}{m}{n}
\rput(1.3745313,-2.4215937){$r(0)$}
\end{pspicture} 
}
\caption{The minimal surface at fixed $y$. We have chosen our coordinates such that $x$ runs from $-L$ to $L$. We use the symmetry $x\to-x$ to reduce the problem such that $x$ runs from $0$ to $L$ instead.}
\label{fig:minimal-surf-strip}
\end{figure}
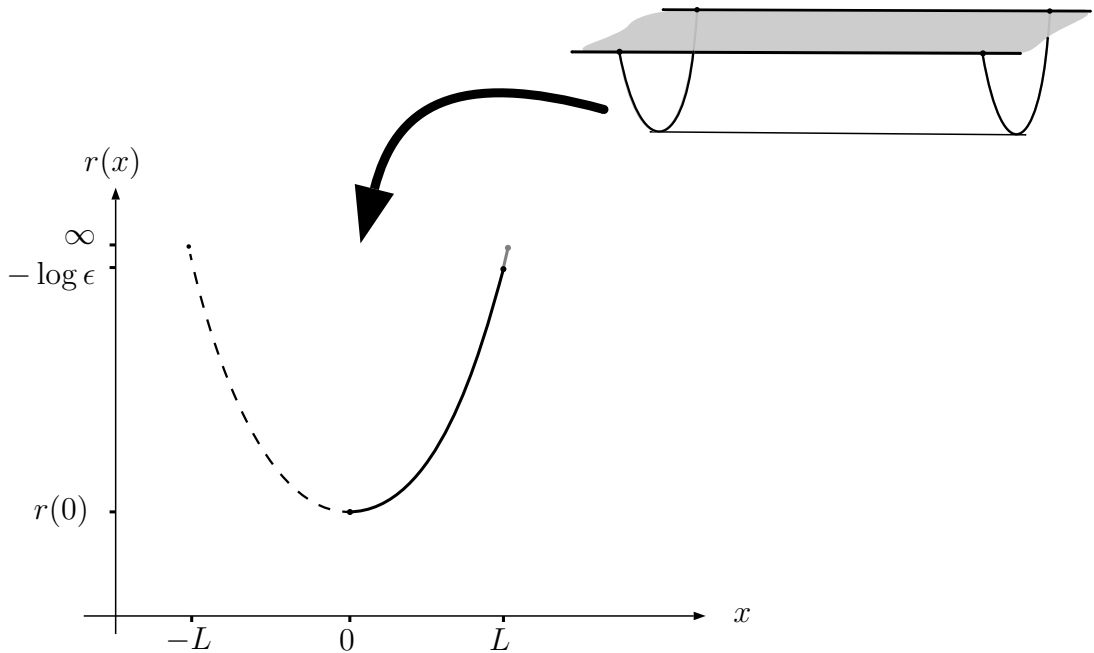

One can associate a Hamiltonian to $x$-evolution,
\begin{align}
H(r,p)\ =\ p\,\dot{r}-\mathcal{L}(r,\dot{r})\,\Big|_{\dot{r}(p,r)}\,,
\end{align}
which is conserved, such that $H(r,p)=E$. The on-shell action is a function of the boundary data $r(0)$ and $r(L)=-\log\epsilon$. The integration constant $r(0)$ is related to the constant of motion $E$ by imposing that $r(0)$ is the turning point of the minimal surface in the bulk, see Figure \ref{fig:minimal-surf-strip}. Thus,
\begin{align}
0\ =\ \dot{r}(0)\ =\ \frac{\partial H}{\partial p}\big(r(0),E\big)\,.
\end{align}
In our case this yields $E=\frac{\widetilde{L}}{2G_d}f\big(r(0)\big)$. So it seems that given $\epsilon$ and $E$, we get a value of the on-shell action. However, the physical input that we give the system is $L$ rather than $E$, so we need to express $E$ in terms of $L$. In summary, the boundary conditions are set by the two integration constants $L$ and $\epsilon$ via:
\begin{align}
f(r(0)) &= E(L)\,,&
r(L) &= \log\frac1\epsilon
\end{align}
Because the Hamiltonian is a constant of motion, we can write the on-shell action as the Legendre transform of the so-called characteristic function $W(\epsilon,E)$,\footnote{See e.g.\ Chapter 10 of \cite{goldstein}.}
\begin{align}\label{eq:on-shell-action-strip}
S_\text{ent}(\epsilon,L)\ &=\ -E\,L+W(\epsilon,E)\,,&
L &= \frac{\partial W}{\partial E}\,.
\end{align}
Now, we assume that the characteristic function is separable, by which we mean that it splits up into two pieces:
\begin{align}\label{eq:separate-W}
W(\epsilon,E)\ =\ W_\epsilon(\epsilon)+W_E(E)\,.
\end{align}
The first piece $W_\epsilon$ contains the area-law divergence, while the second one $W_E$ contains information about the finite piece of the entanglement entropy \eqref{eq:entanglement-strip}. The above separation is justified if there is a clean separation between the UV and IR, which is the case when the entangling surface is a strip \eqref{eq:entanglement-strip}. The reason why we want the separation \eqref{eq:separate-W} is that $L=\frac{\partial W}{\partial E}=W_E'(E)$ depends only on $E$ this way. The renormalized entanglement entropy \eqref{eq:c-function} then becomes simply
\begin{align}\label{eq:holog-c-function}\boxed{
\ c_d(L)\ =\ -\beta_d\,\frac{L^{d-1}}{\widetilde{L}^{d-2}}\ E(L)\ 
}\end{align}
where $E(L)$ is obtained by inverting the relation\footnote{This integral would not converge if we were not allowed to separate the characteristic function as in \eqref{eq:separate-W}. In other words, we would need to introduce the cut-off $\epsilon$. The integral would then run up to $r=1/\epsilon$ instead of all the way to $r\to\infty$, which would introduce a dependence of $E$ on $\epsilon$.}
\begin{align}
L(E)\ =\ \int_{r(0)(E)}^\infty\ \frac{dr}{\dot{r}(r,E)}\ =\ \int_{r(0)(E)}^\infty dr\, \left( \frac{\partial H}{\partial p} \right)^{-1}\,\bigg|_{p=p(E,r)}\,.
\end{align}
We have thus reduced the problem of finding the renormalized entanglement entropy of a strip to inverting $L(E)$ to $E(L)$.

\textbf{A relation between bulk and boundary length scales.}
It is known that the physical scale, i.e.\ the scale at which one probes the theory, is related to a radial scale in the bulk as $\mu\sim \text{e}^{r/\ell}$. Although this relation between bulk and boundary scales formally true, it is not always easy to make this more precise. The holographic version of the renormalized entanglement entropy is a nice quantity to consider, because it gives an \emph{explicit} relation between a boundary scale $\mu=1/L$ and a bulk scale $r(0)$ (or $E$).

\subsection[AdS$_4$ and Lifshitz central charges]{AdS$_{\bf4}$ and Lifshitz central charges}
\begin{figure}[h]\begin{center}
\includegraphics{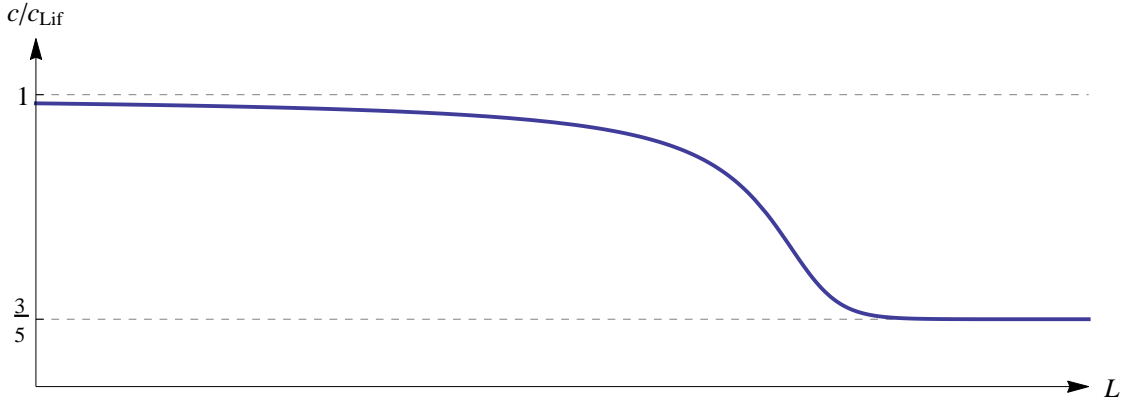}
\caption{The renormalized entanglement entropy is evaluated on the numerical background. Along the horizontal axis we have the width of the strip $L$. We divided by the Lifshitz value $c_\text{Lif}=\frac{\ell^2_\text{Lif}}{2G}$.}
\label{fig:entanglement}
\end{center}\end{figure}

Let us put formula \eqref{eq:holog-c-function} to good use. We restrict ourselves to $d=3$ boundary dimensions henceforth. The AdS$_4$ and Lifshitz backgrounds correspond to $f(r)=\text{e}^{2r/\ell}$, where the curvature length scale is either $\ell=\ell_\text{AdS}$ or $\ell=\ell_\text{Lif}$; they are related as $\ell_\text{AdS}/\ell_\text{Lif}=\sqrt{3/5}$. First of all, we have
\begin{align}
\dot{r}(r,E)\ =\ \text{e}^r\sqrt{\text{e}^{4(r-r(0))/\ell}-1}\,,
\end{align}
where $r(0)$ is given in terms of $E$ via $E=\frac{\widetilde{L}}{4G}\,\text{e}^{2r(0)/\ell}$. Then, we find
\begin{align}
L(E)\ =\ \,\sqrt{\frac{\ell^2}{2G}\,\frac{\widetilde{L}}{\beta_3 E}}\,,
\end{align}
which can easily be inverted to $E(L)$. The central charges for the AdS and Lifshitz fixed points are thus given by
\begin{align}\label{eq:ads-lif-central-charges}
c_\text{AdS}\ &=\ \frac{\ell^2_\text{AdS}}{2G}\,,&
c_\text{Lif}\ &=\ \frac{\ell^2_\text{Lif}}{2G}\,.
\end{align}
As a first consistency check, we see that $c_\text{AdS}/c_\text{Lif}=3/5<1$, which gives credence to the statement that the flow must be from a Lifshitz-type fixed point in the UV to a conformally invariant fixed point in the IR. Using formula \eqref{eq:holog-c-function} it is actually quite easy to evaluate the renormalized entanglement entropy on a numerical background. The result of this is shown in Figure \ref{fig:entanglement}. 
We clearly see that as $L$ becomes large enough, the minimal surface dips into the bulk deep enough to become sensitive to the AdS$_4$ part of the geometry.

\section{Holographic Renormalization}\label{sec:holographic-renormalization}

In this section we show that the on-shell value of the action \eqref{eq:massive-vector-lagrangian} can be renormalized by adding only local covariant counterterms. We will work in the Hamilton--Jacobi formalism \cite{deBoer:1999xf,Martelli:2002sp}, in which the on-shell action is found by solving the Hamiltonian constraint. One usually splits up the on-shell action in terms of a local and a non-local part. When we consider only relevant operators, all power-law divergences are contained in the local part, which consists of a finite number of terms (these will be the counterterms). In the present case, however, we have a marginally relevant operator. For simplicity we assume translational invariance in the boundary directions. In the presence of the marginally relevant operator, we still find that all divergences are contained in the local piece, but we find that we need an infinite number of local counterterms. Because we impose translational invariance, there is only one Lorentz scalar one can construct (namely $A^2$), so the counterterms will just be powers of $A^2$. Although having an infinite number of counterterms may sound problematic, we show that we only really need the first few in order to find the renormalized on-shell action. After that, it is enough to know that one is able to determine all counterterms in principle.

In this section we first compute the counterterm action as an infinite series expansion. Since we have an infinite number of counterterms, the typical way of renormalizing the on-shell action by subtracting counterterms is not very convenient. We propose a new way of finding the renormalized action using the Hamilton--Jacobi equation, which elegantly overcomes the difficulty associated to having an infinite number of counterterms. At the end of this section we will compare this method to the one proposed in \cite{Cheng:2009df}.

\subsection{The Hamilton--Jacobi method}

The generating functional of the field theory is identified with $\log Z$ on the gravity side, which (in the saddle-point approximation) is given by the on-shell value of the action \eqref{eq:massive-vector-lagrangian}. The holographic counterterms can be obtained in the Hamilton--Jacobi (HJ) formalism by assuming that the full on-shell action can be written as the sum of a local piece ($U$) and a non-local piece ($W$).\footnote{The distinction between `local' and `non-local' may seem a bit artificial if one assumes translational invariance. Our working definition of whether something is considered `local' will be whether it is fixed by UV data alone.} We assume translation invariance in the boundary directions, so the most general Ansatz for the on-shell action is a general function of $\alpha\equiv A_aA^a$,
\begin{align} \label{eq:S-split}
I_\text{on-shell}\ =\ \int d^dx\sqrt{g}\,\Big( U(\alpha)+W(\alpha) \Big)\,.
\end{align}
We assume that $U$ can be written as a power series around the Lifshitz point $\alpha=-1$:
\begin{align}\label{eq:U-expansion}
U(\alpha)\ =\ \sum_{n\geq0} u_n(\alpha+1)^n  \,.
\end{align}
The non-local part of the on-shell action $W(\alpha)$ generally cannot be written as a power series. Even though it may not be immediately obvious, we will find that all divergences are contained in $\int d^dx\sqrt{|g|}\, U(\alpha)$, which means that $\int d^dx\sqrt{|g|}\, W(\alpha)$ has a finite limit as $r\to\infty$. The counterterm action is simply
\begin{align}\label{eq:S-ct}
I_\text{ct}\ =\ -\int d^dx\sqrt{g}\, U(\alpha)\,.
\end{align}
The non-local piece will thus be interpreted as the regularized generating functional in the (tentative) dual field theory,
\begin{align}
I_\text{reg}\ =\ \int d^dx\sqrt{g}\, W(\alpha)\,.
\end{align}
The full on-shell action \eqref{eq:S-split} is a solution of the HJ equation 
\begin{align}
\{U+W,U+W\}-\mathcal{L}\ =\ 0\,.
\end{align}
where the brackets are defined as the ``kinetic'' piece of the Hamiltonian constraint, with the momenta replaced by derivatives of the on-shell action e.g.\ $p=\frac{\partial S}{\partial q}$. For the specific theory that we are using as our gravity dual (and assuming translational invariance) we have
\begin{align}
\{F,G\}\ \equiv\ -\int d^dx\sqrt{g}\left[\left(g_{ac}g_{bd}-\frac1{d-1}\,g_{ab}g_{cd}\right)\frac{\partial F}{\partial g_{ab}}\frac{\partial G}{\partial g_{cd}}+\frac12g_{ab}\,\frac{\partial F}{\partial A_a}\frac{\partial G}{\partial A_b}\,\right]\,,
\end{align}
where $g_{ab}$ and $A_a$ are the metric and the vector pulled back onto the constant-$r$ slice. The split of $I_\text{on-shell}$ into a local and a non-local piece induces a split in the HJ equation, 
\begin{align}
\{U,U\}+2\{U,W\}+\{W,W\}-\mathcal{L}(\alpha)\ =\ 0\,.
\end{align}
Our method is the following. First we solve the local part of the HJ equation, 
\begin{align}\label{eq:local-hj-eq}
0\ &=\ \{U,U\}-\mathcal{L}(\alpha) \\
&=\ \frac{3}{8} U^2-\frac{1}{2} \alpha \, U U'-\frac{1}{2} \alpha  (\alpha +4) \left(U'\right)^2+2\alpha -10\,,
\end{align}
which gives us $U(\alpha)$ expanded to arbitrary high order in $(\alpha+1)$. After we have solved \eqref{eq:local-hj-eq}, we plug the solution for $U(\alpha)$ into the non-local part of the HJ equation
\begin{align}
2\{U,W\}+\{W,W\}\ =\ 0\,,
\end{align}
which is solved for the regularized on-shell action $W(\alpha)$.

Finally we should mention that in the case of anomalous breaking of (anisotropic) Weyl symmetry, there are subtleties related to possible mixing of the local with the non-local part of the HJ equation; see \cite{Baggio:2011cp,Baggio:2011ha} for more details. As mentioned before, assuming translation invariance in the boundary directions ensures that such mixing does not occur in this analysis.

\subsection{Covariant counterterms}

The first step in the computation (finding $U(\alpha)$) was done in \cite{Baggio:2011cp}. Let us briefly review this computation. We expand $U(\alpha)$ around the Lifshitz solution ($\alpha=-1$), cf.\ \eqref{eq:U-expansion}. The local part of the HJ equation \eqref{eq:local-hj-eq} can then be solved perturbatively by solving for the coefficients $u_n$ order by order:
\begin{align}
0\, &=\, \{U,U\}-\mathcal{L}(\alpha)
\\\nonumber
&=\,\frac{1}{8} \left(3 u_0^2+4 u_1 u_0+12 u_1^2-96\right)+\left(\frac{u_0 u_1}{4}+u_0 u_2-\frac{u_1^2}{2}+6 u_2 u_1+2 \right)(\alpha+1)+\ldots 
\end{align}
Here and throughout the rest of this section, all ellipses denote $(\alpha+1)\sim1/r$ corrections. This equation has two integration constants that need to be fixed. This is done by fixing the leading-order behavior of the radial Hamiltonian flow, see \S2.3 of \cite{Baggio:2011cp}. The initial conditions come down to setting $u_0=6$ and $u_1=-1$. There is a discrete ambiguity in determining $u_2$, which is related to the choice between the normalizable and the non-normalizable mode for $(\alpha+1)$ (see \S2.4 of \cite{Baggio:2011cp}); one must pick the sign that corresponds to the non-normalizable mode, which gives $u_2=\frac14$. All the other coefficients can then be found recursively. It is straightforward to implement this in an automated algorithm. The first few coefficients are:
\begin{align}
u_0&=6\,,& u_1&=-1\,,& u_2&=\frac14\,,& u_3&=-\frac18\,,& u_4&=\frac{3}{64}\,,
\nonumber\\
u_5&=-\frac1{128}\,,& u_6&=\frac1{256}\,,& u_7&=-\frac{11}{2048}\,,& u_8&=-\frac{9}{4096}\,,& u_9&=\frac{99}{65536}\,.
\end{align}
One can renormalize the on-shell action by simply subtracting $\sqrt{g}\,U(\alpha)$, whose series expansion we just obtained. However, because of the presence of the logarithmic deformation, we need an infinite number of counterterms. So this simple subtraction of counterterms is not the most convenient way to find the renormalized on-shell action. In the present note we take the HJ method one step further and solve for the non-local part of the on-shell action.

\subsection{Renormalized on-shell action}

We will now compute $W(\alpha)$ by solving the non-local part of the HJ equation, $2\{U,W\}+\{W,W\} = 0$. The non-local HJ equation can be rewritten as\footnote{In general, one can see that $\sqrt{g}\,\{U+W,\,\ldots\}=\partial_r(\sqrt{g}\,(\ldots))$, cf.\ \cite{Baggio:2011cp}.}
\begin{align}\label{eq:H_non-local}
\partial_r \left( \sqrt{g}\,W(\alpha) \right)\ =\ \frac12\,\sqrt{g}\,\{W,W\}\,.
\end{align}
We can use the fact that $\partial_r$ generates dilatations (up to $\frac1r$ corrections) to expand $W(\alpha)$ in dilatation weights in the spirit of \cite{Papadimitriou:2004ap},
\begin{align}\label{eq:W-expansion}
W(\alpha)\ &=\ \sum_{n\geq n_0}W_n(\alpha)\,,
&
\frac{\partial_rW_{n}}{W_{n}}\ &=\ -n+O(r^{-1})\,,
\end{align}
such that \eqref{eq:H_non-local} becomes a set of decent equations:
\begin{align}
2(4-n)\,W_{n}\ =\ \sum_{j+k=n}\{W_{j},W_{k}\}\,.
\end{align}
The above expansion simplifies our analysis, because the right-hand side of \eqref{eq:H_non-local} vanishes at lowest order, which fixes the starting point $n_0$ of the expansion to be $n_0=4$. For this it also follows that the non-trivial terms in the expansion \eqref{eq:W-expansion} are given by $W_{4k}$ with $k=1,2,3,...$\,. Let us switch back to how the non-local part of the HJ was written before. The expansion \eqref{eq:W-expansion} gives
\begin{align}
2\{U,W_n\}\ =\ -\sum_{j+k=n}\{W_j,W_k\}\,.
\end{align}
The leading order ($n=4$) part can then be written as
\begin{align}\label{eq:H_marg-expanded}
\varphi(\alpha)\,W_4(\alpha)\ =\ \beta(\alpha)\,W_4'(\alpha)\,,
\end{align}
where $\varphi(\alpha)$ and $\beta(\alpha)$ are known functions of $U(\alpha)$ and $U'(\alpha)$:
\begin{align}
\varphi(\alpha)\ &\equiv\   -\frac34 U(\alpha )+\frac12 \alpha \, U'(\alpha )\,, 
\\
\beta(\alpha)\ &\equiv\   -\frac12\alpha\,U(\alpha )- (\alpha^2 +4\alpha)\, U'(\alpha )\,.
\end{align}
Remember that at this stage in the calculation the power-series expansion for $U(\alpha)$ is known up to arbitrarily high order. The solution of the marginal piece of the HJ equation is then
\begin{align}\label{eq:W-sol}
W_4(\alpha)\ =\ \exp\left[\int d\alpha\,\frac{\varphi(\alpha)}{\beta(\alpha)}\right]\,,
\end{align}
The function $W_4(\alpha)$ is non-analytic in some point $\alpha=\alpha_0$ if the integrand $\varphi(\alpha)/\beta(\alpha)$ has a pole there.\footnote{This is not strictly true. For instance, let us say $\varphi(\alpha)/\beta(\alpha)=c/(\alpha-\alpha_0)$ then $W(\alpha)=(\alpha-\alpha_0)^c$, which is only non-analytic in $\alpha=\alpha_0$ if $c<1$.} Moreover, the function $\beta(\alpha)$ only has a zero when there is some sort of scale invariance. Namely, there is a reason we use the notation $\beta(\alpha)$, since this function can be interpreted as the \emph{beta-function} of $\alpha$, that is $\beta(\alpha)=\partial_r\alpha$ up to $1/r$ corrections, cf.\ \cite{Baggio:2011cp}.\footnote{The physical mass scale $\mu$ in standard AdS/CFT is identified with the radial coordinate as $\mu\leftrightarrow\text{e}^r$, which is why $\mu \partial_\mu\leftrightarrow\partial_r$. Also, there is a relation similar to $\beta(\alpha)=\partial_r\alpha$ for the other function, $\varphi(\alpha)$. Namely the function $\varphi(\alpha)$ can be seen as the ``beta function'' associated to the scaling of the volume form, because it turns out that $\varphi(\alpha)=\partial_r\log(\sqrt{|g|})$.} So, a zero in $\beta(\alpha)$ means that there is some scale-invariant fixed point at which the renormalized on-shell action is non-analytic.

Let us return to the calculation. The series expansion of $U(\alpha)$ induces a series expansion for the integrand in \eqref{eq:W-sol}:
\begin{align}\label{eq:phi/beta}
\frac{\varphi(\alpha)}{\beta(\alpha)}\ =\ \frac8{(\alpha+1)^2}+\frac{10}{\alpha+1}+\ldots\,,
\end{align}
such that\footnote{This expression for $W_4$ also appeared in \cite{marco-thesis}.}
\begin{align}
W_4(\alpha)\ &=\ w\,\varepsilon(\alpha)+\ldots\,,
&
\varepsilon(\alpha)\ &\equiv\ \text{e}^{-\frac8{\alpha+1}}\,(\alpha+1)^{10}\,.
\end{align}
where $w$ is some integration constant that comes from the $\alpha$ integral and we defined $\varepsilon(\alpha)$, because it provides us with a nice expansion parameter. The non-analytic nature (around $\alpha=-1$) of this function is obvious from the factor $\text{e}^{-8/(\alpha+1)}$. The result of the expansion in powers of $\varepsilon(\alpha)$ is
\begin{align}\label{eq:W-coefficients}
W_{4k}(\alpha)\ =\frac{\tilde{w}_k\,(\varepsilon w)^{k}}{(\alpha+1)^{4(k-1)}}+  \ldots
\end{align}
where the first few coefficients are
\begin{align}
\tilde{w}_0&=1\,,
&
\tilde{w}_1&=24\,,
&
\tilde{w}_2&=1152\,,
&
\tilde{w}_3&=73728\,,
&
\tilde{w}_4&=5529600\,.
\end{align}
We can check explicitly whether the renormalized on-shell action is finite by evaluating it on the asymptotic solution from \cite{Cheng:2009df}, cf. Appendix \ref{sec:asymptotic-solution},\footnote{In order to have the source $\Lambda$ appear as a coefficient in the expression for $\alpha$ we used \eqref{eq:scaling-transformation}, such that we expand in $r$ rather than $r-\log\Lambda$.}
\begin{align}\label{eq:alpha-asympt}
\alpha_\text{asymp}\ &=\ -1+\frac{2}{r}+\frac{2\log\Lambda+5 \log r+2-\lambda}{r^2}+\ldots\,,
\\
\varepsilon(\alpha_\text{asymp})\ &=\ \text{e}^{-4r} \left(\Lambda^4\,\text{e}^{4-2\lambda}+\ldots\,\right)\,.
\end{align}
It is pleasing to see that the somewhat convoluted expressions for $\alpha_\text{asymp}$ and $\varepsilon(\alpha)$ conspire to give this simple power-law form. The integration constant $\Lambda$ may be seen as the source for the marginal operator. Besides $\Lambda$ we also have the metric sources:
\begin{align}\label{eq:metric-sources}
N\ &\equiv\ \lim_{r\to\infty}\text{e}^{-2r}r^2\,(-g_{tt})^{1/2}\,,
&
h\,\delta_{ij}\ &\equiv\ \lim_{r\to\infty}\text{e}^{-2r}r^{-2}\,g_{ij}\,.
\end{align}
The renormalized on-shell action can be written in terms of these sources as
\begin{align}
I_\text{ren}\ =\ \lim_{r\to\infty}\,\sqrt{g}\,W(\alpha)\ =\ Nh\,\Lambda^4\,\text{e}^{4-2\lambda}\,w\,.
\end{align}

\subsection{One-point functions}

The one-point functions can be computed by taking derivatives with respect to the sources. In \cite{Ross:2011gu} it was argued that the appropriate sources are frame fields rather than the metric components. That is, the vector field cannot be viewed separately from the metric, because both depend on the frames: $g_{ab}=\eta_{\scriptscriptstyle AB}\,e^{\scriptscriptstyle A}_ae^{\scriptscriptstyle B}_b$ and $A_a=A_{\scriptscriptstyle A}\,e^{\scriptscriptstyle A}_a$. The frames are chosen such that the vector lies along the $0$-direction in the tangent space: $A_a=\sqrt{-\alpha}\,e^0_a$. The degrees of freedom in this description are thus $e^{\scriptscriptstyle A}_a$ and $(\alpha+1)$. The corresponding conserved charges are\footnote{In the metric formalism, the HJ momenta are given by the derivatives of the on-shell action (characteristic function) with respect to the fields, i.e.\ $\sqrt{g}\,\pi^{ab}=\delta I_\text{on-shell}/\delta g_{ab}$ and $\sqrt{g}\,E^a=\delta I_\text{on-shell}/\delta A_a$.}
\begin{align}\label{eq:Tab}
\mathcal{T}^a{}_b\ &\equiv\ \frac{1}{\sqrt{g}}\,e^{\scriptscriptstyle A}_b\,\frac{\delta I_\text{on-shell}}{\delta e^{\scriptscriptstyle A}_a}\ =\ 2\pi^a{}_b+E^aA_b
\\
\pi_\alpha\ &\equiv\ \frac1{\sqrt{g}}\,\frac{\delta I_\text{on-shell}}{\delta \alpha}
\end{align}
The non-zero (bare) charges in our Ansatz are $\pi_\alpha$, the energy density $\mathcal{E}=\mathcal{T}^t{}_t$ and the pressure density $\mathcal{P}=-\mathcal{T}^x{}_x=-\mathcal{T}^y{}_y$. A Weyl transformation acts on the fields as (see also \cite{Ross:2011gu,Baggio:2011cp})
\begin{align}
\delta_\omega e^0_a\ &=\ z\omega\,e^0_a\,,
&
\delta_\omega e^{\scriptscriptstyle I}_a\ &=\ \omega\,e^{\scriptscriptstyle I}_a\,,
&
\delta_\omega(\alpha+1)\ &=\ \lambda_\alpha\omega\,(\alpha+1)\,,
\end{align}
with $\lambda_\alpha\equiv -\frac12\left(z+2-\sqrt{9z^2-20z+20}  \right)$. The (bare) on-shell action thus transforms as
\begin{align}
\delta_\omega I_\text{on-shell}\ =\ \int d^dx\sqrt{g}\,\left( z \mathcal{E}-2\mathcal{P}+\lambda_\alpha \pi_\alpha \right)\,\omega\,,
\end{align}
We see that $\lambda_\alpha$ vanishes for $z=2$ in our case. This means that Lifshitz scaling symmetry is preserved if the equation of state $\mathcal{E}=\mathcal{P}$ is satisfied. We investigate whether this Lifshitz equation of state is satisfied in our present setup below.

The momenta that appear in \eqref{eq:Tab} split up due to the split \eqref{eq:S-split}, such that $\pi^{ab}=\pi^{ab}_{\scriptscriptstyle U}+\pi^{ab}_{\scriptscriptstyle W}$ and $E^{a}=E^{a}_{\scriptscriptstyle U}+E^{a}_{\scriptscriptstyle W}$, where
\begin{align}\label{eq:W-momenta}
\pi^{ab}_{\scriptscriptstyle W}\ &=\ \frac12g^{ab}\,W(\alpha)-A^aA^b\,W'(\alpha)\,,
&
E^a_{\scriptscriptstyle W}\ &=\ 2A^a\,W'(\alpha)\,,
\end{align}
and similarly for $W\to U$. This gives the simple expression for the tensor \eqref{eq:Tab}, such that $\mathcal{T}_{\scriptscriptstyle W}^a{}_b=W(\alpha)\,\delta^a_b$. In order to be able to take derivatives with respect to the sources, one rescales the fields in such a way that the rescaled fields have a finite limit as $r\to\infty$, so
\begin{align}
\hat{N}\ &\equiv\ \text{e}^{-2r}r^2\,e^0_t\,,
&
\hat{h}\,\delta_{ij}\ &\equiv\ \text{e}^{-2r}r^{-2}\,\eta_{\scriptscriptstyle AB}\,e^{\scriptscriptstyle A}_ie^{\scriptscriptstyle B}_j\,,
&
\hat{\alpha}\ &\equiv\ \alpha\,.
\end{align}
These rescaled fields reduce to $N$ and $h$ from \eqref{eq:metric-sources} in the limit $r\to\infty$. Notice that $\alpha$ does not change under these field rescalings. The regularized on-shell action becomes:
\begin{align}
\sqrt{g}\,W(\alpha)\ &=\ \hat{N}\hat{h}\, \hat{W}(\alpha)\,,
\end{align}
where we introduced $\hat{W}\equiv \text{e}^{4r}\,W$. The renormalized energy and pressure one-point functions are thus given by the simple expressions
\begin{align}\label{eq:E-and-P-ren}
\hat{\mathcal{E}}\ &=\ \hat{W}(\alpha)\,,
&
\hat{\mathcal{P}}\ &=\ -\hat{W}(\alpha)\,,
\end{align}
Before we can evaluate these on the asymptotic solution, we need to figure out what how $w$ is related to the integration constants of the solution.

\subsubsection*{Finding \textit{w}.}
Since $W(\alpha)=\varepsilon(\alpha)\,w$ up to $1/r$ corrections, we see that we can extract $w$ by taking the derivative of $W(\alpha)$ with respect to $\varepsilon(\alpha)$:
\begin{align}\label{eq:response}
w\ =\ \frac{\partial W}{\partial \varepsilon}\ &=\ \frac{\partial \alpha}{\partial \varepsilon}\,\frac{\partial W}{\partial\alpha}\ =\ \frac{(\alpha+1)^2}{2 \,(5\alpha+9)}\,\frac{W'(\alpha)}{\varepsilon(\alpha)}\,.
\end{align}
The next step is to find an expression for $W'(\alpha)$. Consider the canonical (radial) momentum conjugate to the vector $A_a$, which is related to the vector field as $E_a=-\partial_rA_a$. When we contract the momentum $E^a$ with $A_a/2\alpha$ and use \eqref{eq:W-momenta}, we find the relation
\begin{align}
W'(\alpha)\ &=\ -U'(\alpha)-\frac1{2\alpha}\,A^a\,\partial_rA_a\,.
\end{align}
We can plug this into \eqref{eq:response}, so that we are left the following covariant expression for $w$:
\begin{align}
w\ =\ -\lim_{r\to\infty}\,\frac{\text{e}^{\frac8{\alpha+1}}(\alpha+1)^{-8}}{2 \,(5\alpha+9)}\,\left( U'(\alpha)+\frac1{2\alpha}\,A^a\,\partial_rA_a \right)\,.
\end{align}
When we evaluate this on the asymptotic solution listed in Appendix \ref{sec:asymptotic-solution}, we get:
\begin{align}
w\ =\ \text{e}^{2\lambda-4}\mathcal{M}\,,
\end{align}
where $\mathcal{M}$ is the integration constant that is associated to the normalizable mode as expected, cf.\ Appendix \ref{sec:asymptotic-solution}. We have thus obtained a finite expression for the on-shell action, given in terms of the integration constants of the asymptotic solution:
\begin{align}\label{eq:I_ren}\boxed{
\phantom{\frac11} I_\text{ren}\ =\ Nh\,\Lambda^4 \mathcal{M} \phantom{\frac11}
}\end{align}
The renormalized energy and pressure densities \eqref{eq:E-and-P-ren} associated to the asymptotic solution are $\hat{\mathcal{E}}=-\hat{\mathcal{P}}=\Lambda^4 \mathcal{M}$, which is the same as what one would obtain by taking derivatives of \eqref{eq:I_ren} with respect to $N$ and $h$, switching off the sources ($N=h=1$) afterwards.

Notice that the Lifshitz equation of state is \emph{not} satisfied for non-vanishing $\Lambda$ and $\mathcal{M}$:\footnote{The right-hand side of this equation is non-zero even when $\Lambda=0$ if we consider non-translationally invariant configurations \cite{Griffin:2011xs,Baggio:2011ha,Chemissany:2012du}.}
\begin{align}\label{eq:tracelessness-condition}
\hat{\mathcal{E}}-\hat{\mathcal{P}}\ =\ 2\Lambda^4\mathcal{M}\,,
\end{align}
As was emphasized in \cite{Cheng:2009df}, $\Lambda$ should be regarded as a dynamically generated scale.\footnote{The pure Lifshitz spacetime geometry can be obtained from the asymptotic solution by taking $\Lambda\to0$ while holding $r$ fixed and rescaling $t$ and $x$ accordingly.} In other words, Lifshitz scaling symmetry is broken dynamically.

Finally, notice that the renormalized free energy does not depend on $\lambda$ (or $\tilde\lambda$). This is somewhat curious, because in principle one could expect any function of the dimensionless\footnote{By `dimensionless' we mean invariant under the Lifshitz scaling transformation \eqref{eq:scaling-transformation}.} combinations $\Lambda^4 \mathcal{M}$ as well as $\Lambda^2\,\text{e}^{-\lambda}$. At this stage, we do not have an intuitive understanding of why this is the case. It might be similar to what we know from planar $\mathcal{N}=4$ SYM, where the free energy is independent of the marginal (gauge) coupling at weak and at strong coupling, though the exact answer does depend on it.

\subsubsection*{HJ method for numerics.}

\begin{figure}[h]\begin{center}
\includegraphics{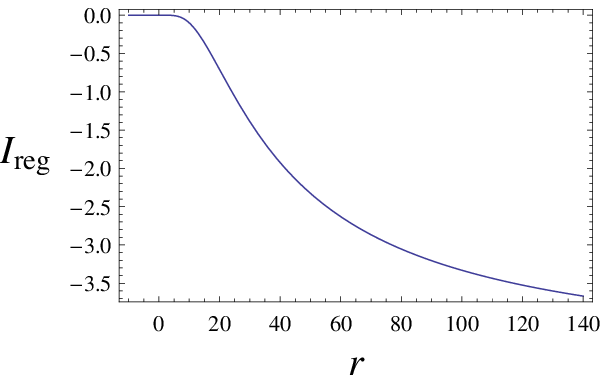}
\qquad\qquad
\includegraphics{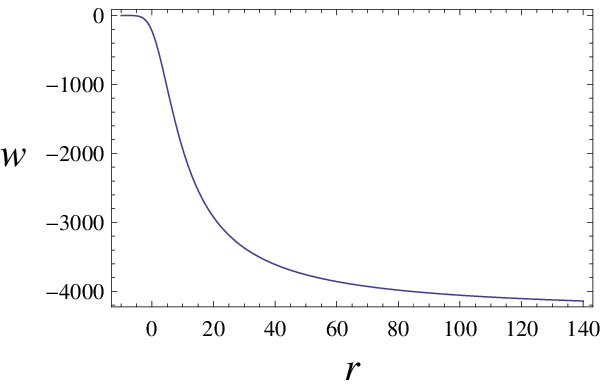}
\caption{The regularized on-shell action $I_\text{reg}$ and the expectation value $w$ are evaluated on the numerical interpolating solution. Although $I_\text{reg}$ tends to a constant only logarithmically, its asymptotic value can be found by fitting it against $a+\frac{b}{r}+\frac{c+d\,\log r}{r^2}$ to extract $a$. In this case we find $I_\text{ren}=\lim_{r\to\infty}I_\text{reg}\approx-4.67$.}
\label{fig:I_reg} 
\end{center}\end{figure}

Now that we expressed the expectation value in terms of the normalizable mode $\beta$, we could (in principle) compare it to the value of beta of the numerical interpolating solution. However, the usual method of inverting the asymptotic series is not easy, because the power-law expansion is contaminated by the logs. Another way to find the renormalized on-shell action $I_\text{ren}$ and the expectation value $w$ is to integrate $2\{U,W\}+\{W,W\}=0$ numerically. This is quite straightforward to do, since it is just a first-order equation albeit non-linear. The only thing one should be careful of is not to expand $U(\alpha)$ to arbitrarily high order, because it is an asymptotic series \cite{marco-thesis}. The result of this is shown in Figure \ref{fig:I_reg}.

\subsubsection*{Comparison with previous result}

Let us compare our results to \cite{Cheng:2009df}. The method of renormalizing the free energy used in \cite{Cheng:2009df} does not follow the typical prescription of removing divergences by adding local covariant variant counterterms. The proposed counterterms were written on a local covariant basis,  $c_0\,\sqrt{g}$, $c_1\,\sqrt{g}\,(\alpha+1)$ and $c_2\,\sqrt{g}\,(\alpha+1)^2$, but their coefficients were allowed to be explicit functions of the cut-off of the form $c_n\sim 1+1/r + 1/r^2+\ldots$. Because the dependence of the renormalized on-shell action on $\alpha=A^2$ is different among the two methods, the one-point functions obtained through functional derivation is also different. Indeed, we find that the expressions for the one-point functions do \emph{not} agree.

Another mismatch between this work and \cite{Cheng:2009df} comes from the integration constants. Looking at the Euler--Lagrange equations (Eqs.\ (14)--(16) of  \cite{Cheng:2009df}), one expects to find five integration constants. The generic asymptotic solution that one finds, however, has six. One of the integration constants must thus be fixed. The HJ formalism requires that this spurious integration constant be fixed in a very specific way, which does not break Lifshitz symmetry explicitly. As it turns out, the way the integration constant was fixed in \cite{Cheng:2009df} does break Lifshitz symmetry explicitly. See Appendix \ref{sec:asymptotic-solution} for more details.

\section{Conclusion}

We have seen that the leading logarithmic modes that are present in the solution of the field equations in the massive-vector model can be interpreted as a marginally relevant operator in the $z=2$ Lifshitz field theory. The theory flows to a conformally invariant fixed point in the IR due to this marginally relevant operator. We derived a nice formula for the holographic entanglement $c$-function and we saw how it decreased along the RG flow. We found a way to renormalize the on-shell action without introducing explicit dependence on the radial cutoff. This did require that we have an infinite number of counterterms, although we showed that one can obtain the renormalized on-shell action without knowing all counterterms explicitly.

The Hamilton--Jacobi method typically has an ambiguity that comes from the freedom to add finite local counterterms to the renormalized on-shell action. This ambiguity is absent in our case, because there are no finite local covariant counterterms when one assumes translational invariance in the boundary directions. This can also be seen in the free 3D Lifshitz scalar, $\dot\phi^2+(\nabla^2\phi)^2$, where there are no finite counterterms at the non-derivative level.

The closed-form expression we found for the renormalized on-shell action is a non-analytic function of $A^2+1$, which reduces to a very simple expression once it is evaluated on the asymptotic solution. We computed the expectation values if the renormalized energy and pressure densities and we found that the Lifshitz equation of state $z\mathcal{E}=d_s\mathcal{P}$ (which is equivalent to the anisotropic tracelessness condition) is broken dynamically.

A possible extension of this work would be to study the thermodynamic properties of these asymptotically Lifshitz geometries. In particular, it would be interesting to find an AdS-to-Lifshitz crossover in the free energy as a function of temperature. The temperature-scaling of the free energy depends on the dynamical exponent $z$, $\mathcal{F}\sim T^{1+d_s/z}$, so one may see a transition $\mathcal{F}\sim T^3$ to $\mathcal{F}\sim T^2$ as $T$ is increased. Another extension of this work would be to let go of translation invariance, though at this stage this seems to complicate matters quite severely. 

The Hamilton--Jacobi method proved powerful when dealing with the presence a marginal operator. It would be interesting to see how this method applies to the more canonical example of Einstein gravity coupled to a marginal scalar.

\section*{Acknowledgements}

I am grateful to Marco Baggio and Jan de Boer, with whom a substantial part of this work was done in collaboration.\footnote{Part of this work appeared in Marco Baggio's PhD thesis \cite{marco-thesis}.} I would also like to thank Sean Hartnoll, Jelle Hartong, Micha\l\ Heller, Diego Hofman, Paul de Lange, Matthew Lippert and Blaise Rollier for useful discussions. Furthermore I would like to acknowledge Bram Wouters for his contributions in the early stages of this project. This work is part of the research program of the Foundation for Fundamental Research on Matter (FOM), which is part of the Netherlands Organization for Scientiﬁc Research (NWO).

\newpage
\appendix

\section{Asymptotic solution}\label{sec:asymptotic-solution}

The asymptotic solution was found in \cite{Cheng:2009df}. Because the solution itself does not look too pretty we have kept it out of the main text. The non-zero metric and vector components are
\begin{align}
\nonumber\\
g_{tt}\ &=\ -N^2\,\frac{\Lambda^4\,\text{e}^{4\rho}}{\rho^4}\,\left( 1+\frac{10 \log\rho+10-2 \lambda}{\rho}+\ldots \right)
\nonumber\\
&\qquad+\frac{\mathcal{M}}{4\rho^2}\,\left(1+\frac{ 5  \log\rho+\tfrac{41}{6}-2 \lambda-\tilde{\lambda } }{\rho}+\ldots \right)
\\\nonumber\\
g_{ii}\ &=\ h\,\Lambda^2\,\rho^2\,\text{e}^{2\rho}\,\left( 1+\frac{5 \log\rho+4-\lambda}{\rho}+\ldots \right)
\nonumber\\
&\qquad+\frac{\mathcal{M}\,\rho^4\,\text{e}^{-2\rho}}{8}\,\left(-1+ \frac{10\log\rho+\tfrac{13}{6} - \lambda + \tilde{\lambda } }{\rho }+\ldots \right)
\\\nonumber\\
A_t\ &=\ N\,\frac{\Lambda^2\,\text{e}^{2\rho}}{\rho^2}\,\left( 1+\frac{5 \log\rho+4-\lambda}{\rho}+\ldots \right)
\nonumber\\
&\qquad+\frac{5\mathcal{M}\,\text{e}^{-2\rho}}{8}\,\left( 1-\frac{\tilde{\lambda }+\lambda -\tfrac{35}{6}}{\rho } +\ldots\right)
\\\nonumber
\end{align}
The ellipses denote terms that are sub-leading in $(\log\rho)/\rho$. The radial coordinate we use here is related to the one in  \cite{Cheng:2009df} as $\rho=-\log(\Lambda r_\text{\cite{Cheng:2009df}})$. In terms of the radial coordinate $r$ that we use throughout the rest of this note, we have $\rho=r-\log\Lambda$. One can see that the pure Lifshitz geometry is obtained by $\Lambda\to0$ keeping $r$ fixed (and rescaling $t$ and $\vec{x}$ accordingly). The integration constants $(N,h,\lambda,\mathcal{M},\tilde{\lambda})$ are related to the ones in \cite{Cheng:2009df} as
\begin{align}
N &= \sqrt{f_0}\,,
&
h &= p_0\,,
&
\lambda &= \lambda\,,
&
\mathcal{M} &= \frac{4\beta}{3\sqrt{2}}\,,
&
\tilde{\lambda} &= \frac{\alpha}{\beta}\,,
\end{align}
where $\alpha$ is an integration constant that appears in \cite{Cheng:2009df}, it is \emph{not} $A^2$. A useful contraction that we use in the main text is:
\begin{align}
\alpha\ \equiv\ A^aA_a\ &=\ -1+\frac{5 \log\rho+2-\lambda}{\rho ^2}+\frac{2}{\rho }+\ldots
\nonumber\\
&\qquad+\frac{3\mathcal{M}\,\rho^2\,\text{e}^{4\rho}}{2}\,\left( -1 + \frac{5\log(\rho)+1+\tilde{\lambda}}{\rho} \right)
\end{align}
A Lifshitz scaling transformation acts on the integration constants in the following way:
\begin{align}\label{eq:scaling-transformation}
\left( \Lambda, \lambda, \mathcal{M}, \tilde{\lambda} \right)\ \to\
\left( \text{e}^{\lambda'/2}\Lambda, \lambda+\lambda', \text{e}^{-2\lambda'}\mathcal{M}, \tilde{\lambda}-\lambda' \right)
\end{align}
So a Lifshitz rescaling can be seen as a redefinition of the scale $\Lambda$. In the gauge chosen both here as well as in \cite{Cheng:2009df}, there is one spurious integration constant, which must be removed from the solution. This spurious integration constant should be removed by fixing a dimensionless combination of integration constants, which is a combination that is invariant under \eqref{eq:scaling-transformation}. This is necessary so as not to break the Lifshitz symmetry explicitly. The Hamilton--Jacobi formalism will tell us uniquely which dimensionless combination we must fix. In \cite{Cheng:2009df} the extra integration constant is removed in a way that does not preserve \eqref{eq:scaling-transformation}, thereby breaking Lifshitz symmetry \emph{explicitly}. 

We will now show how the spurious integration constant is fixed in the HJ formalism. Consider the canonical momenta 
\begin{align}
\pi^{ab}\ &=\ \frac{1}{\sqrt{g}}\,\frac{\delta I_\text{on-shell}}{\delta g_{ab}}
&
E^a\ &=\ \frac{1}{\sqrt{g}}\,\frac{\delta I_\text{on-shell}}{\delta A_a}
\end{align}
which split up into $\pi^{ab}=\pi^{ab}_{\scriptscriptstyle U}+\pi^{ab}_{\scriptscriptstyle W}$ and $E^{a}=E^{a}_{\scriptscriptstyle U}+E^{a}_{\scriptscriptstyle W}$, where
\begin{align}
\pi^{ab}_{\scriptscriptstyle W}\ &=\ \frac12g^{ab}\,W(\alpha)-A^aA^b\,W'(\alpha)
&
E^a_{\scriptscriptstyle W}\ &=\ 2A^a\,W'(\alpha)
\end{align}
and similarly for $W\to U$. It is useful to define the tensor
\begin{align}\label{eq:noether-T}
T^{ab}\ \equiv\ 2\pi^{ab} + E^{(a}A^{b)}\ =\ g^{ab} \left( U+W \right)
\end{align}
From these expressions we can isolate $W(\alpha)$ and $W'(\alpha)$ by respectively taking the trace of $T^{ab}$ and by contracting $E^a$ with $A_a/2\alpha$:
\begin{align}\label{eq:W-U}
W(\alpha)\ &=\ \frac13\,\left( 2g^{ab}\,\partial_rg_{ab}-A^a\,\partial_rA_a \right) -U(\alpha)
\\\label{eq:W'-U'}
W'(\alpha)\ &=\ -\frac1{2\alpha}\,A^a\,\partial_rA_a -U'(\alpha)
\end{align}
where we used the canonical relations $\pi_{ab}=-K_{ab}+g_{ab}\,K$ (with $K_{ab}=\frac12 \partial_rg_{ab}$) and $E_a=\partial_rA_a$. On the other hand, from the leading-order Hamilton--Jacobi equation for $W$ one finds
\begin{align}
\sqrt{g}\,W(\alpha)\ =\ \frac{(\alpha+1)^2}{2(5\alpha+9)}\,\sqrt{g}\,W'(\alpha)+\ldots
\end{align}
where the ellipses denote $\alpha+1\sim 1/\rho$ corrections. Thus, if one computes the renormalized on-shell action on the asymptotic solution using \eqref{eq:W-U} and \eqref{eq:W'-U'} one should get the same answer on both sides of the equation. On the left-hand side we get (up to $1/r$ corrections)\footnote{Subtracting $U$ in \eqref{eq:W-U} ensures that all power-law ($\sim \text{e}^{4\rho}$), logarithmic ($\sim \rho^\#$), and double-logarithmic ($\sim \log\rho$) divergences cancel. This can be checked explicitly up to arbitrarily high order in the asymptotic expansion.}
\begin{align}
\sqrt{g}\,W(\alpha)\ =\ -\frac{1}{3} Nh\, \mathcal{M}\, (\lambda +\tilde{\lambda}-\frac{17}{6})
\end{align}
while on the right-hand side we get
\begin{align}
\frac{(\alpha+1)^2}{2(5\alpha+9)}\,\sqrt{g}\,W'(\alpha)\ =\ -Nh\, \mathcal{M}
\end{align}
Comparing these two expressions gives
\begin{align}\label{eq:lambda-fixed}
\lambda+\tilde{\lambda}\ =\ -\frac16\,,
\end{align}
Notice that this does not affect the scaling transformation \eqref{eq:scaling-transformation}, because the combination $\lambda+\tilde{\lambda}$ is invariant. A further check that this is the correct value for $\lambda+\tilde{\lambda}$ comes from looking at the expectation value of the energy and pressure densities, which turn out to be related to the tensor \eqref{eq:noether-T}:
\begin{align}
\langle \mathcal{E}\rangle\ &=\ \lim_{\rho\to\infty} \text{e}^{4\rho}\, T_{\scriptscriptstyle W}^t{}_t\ =\ \mathcal{M}\,\left(\lambda  +\tilde{\lambda} +\frac{7}{6}\right)\,,
\\
\langle \mathcal{P}\rangle\ &=\ \lim_{\rho\to\infty} \text{e}^{4\rho}\, T_{\scriptscriptstyle W}^x{}_x\ =\ -\mathcal{M}\,\left( \lambda + \tilde\lambda-\frac{5}{6} \right)\,,
\end{align}
where we used the same trick with the canonical momenta. For the special value \eqref{eq:lambda-fixed} we get
\begin{align}
\langle \mathcal{E}\rangle\ &=\ \mathcal{M}\,,
&
\langle \mathcal{P}\rangle\ &=\ \mathcal{M}\,.
\end{align}
We thus see that the equation of state $\langle \mathcal{E}\rangle=\langle \mathcal{P}\rangle$ holds, i.e.\ $2T_{\scriptscriptstyle W}^t{}_t=T_{\scriptscriptstyle W}^i{}_i$. The fact that this had to be true can be seen directly from \eqref{eq:noether-T}:
\begin{align}
T_{\scriptscriptstyle W}^a{}_b\ =\ \delta^a_b\, W(\alpha)
\end{align}
Finally, we should mention that one can expand in $r$ rather than $\rho=r-\log\Lambda$. This comes down to performing a rescaling \eqref{eq:scaling-transformation} with $\lambda'=-2\log\Lambda$. This is the convention we use in the main text.

\section{Numerics}\label{sec:numerics}

In this section we set up the numerical solution that interpolates between AdS in the interior and Lifshitz in the asymptotic region, see e.g.\ \cite{Braviner:2011kz} for previous work on flows that involve a Lifshitz scaling region in the massive-vector model. We use the Ansatz consistent with translational invariance and we focus on scalar modes only. The Ansatz is:
\begin{align}
ds^2\ &=\ -f(r)\,dt^2+g(r) \left( dx^2+dy^2 \right)+dr^2\,,
&
A\ &=\ h(r)\,dt\,.
\end{align}

\begin{figure}[h!]\begin{center}
\includegraphics{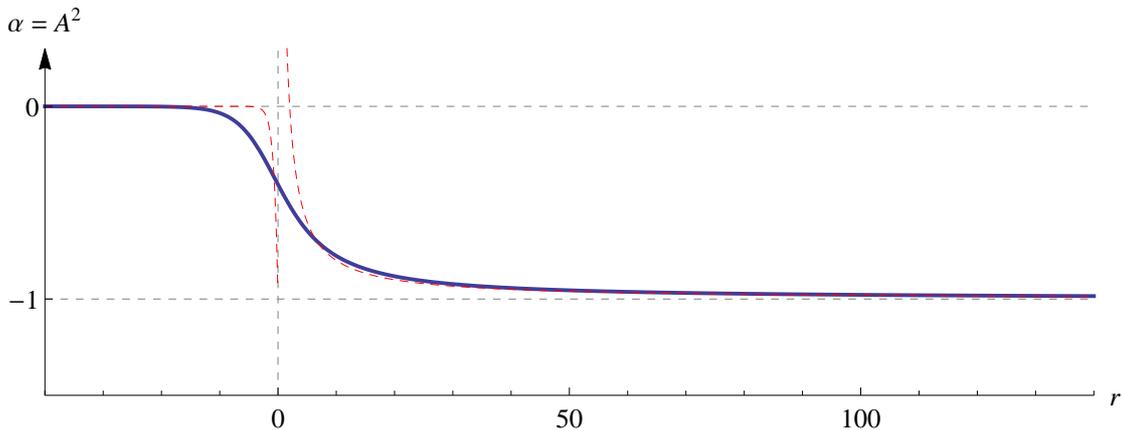}
\caption{The quantity $\alpha=A_aA^a$ is evaluated on the numerical solution. On the left (IR) we have AdS, $\alpha=0$, and the right (UV) we have log-Lifshitz, $\alpha=-1$. The red dashed curves are the approximate analytic solutions $\alpha=\text{e}^{\Delta(r-r_*)}$ with $\Delta=2(\nu-1)/\ell$ (left) and $\alpha=-1+\frac2{r-r_*}$ (right).}
\label{fig:alpha-numerics-1}
\end{center}\end{figure}

\textbf{Shooting method.}
We will use the shooting method where we shoot from the AdS solution outward. The AdS background is (absorbing $\ell$ factors into $t$, $x$, $y$, cf.\ \eqref{eq:ads-metric})
\begin{align}
f(r)\ &=\ \text{e}^{2r/\ell}\,,
&
g(r)\ &=\ \text{e}^{2r/\ell}\,,
&
h(r)\ &=\ 0\,.
\end{align}
We work in coordinates such that the Lifshitz curvature scale is set to one, which fixes the AdS scale to $\ell=\sqrt{3/5}$. In order to ensure that the solution flows to Lifshitz quickly enough we turn on the source for the irrelevant operator discussed in the main text. The linearized mode that plays the role of this source is
\begin{align}
\delta h(r)\ &=\ \varepsilon\,\text{e}^{\nu r/\ell}\,,
&
\nu\ &=\ \frac12 \left( -1+\sqrt{1+16\ell^2} \right)\,.
\end{align}
The small parameter $\varepsilon$ sets the radial scale at which the irrelevant mode picks up speed. To be more precise, the crossover point is at $r\sim r_*$, where
\begin{align}\label{eq:crossover-point}
r_*=\frac\ell\nu\,\log\left(1/\varepsilon\right)\,.
\end{align}
We will let the numerical integration run from $r=-40$ to $r=140$, and we set the crossover radius to zero, so $\varepsilon=1$. The result of this calculation is plotted in terms of $\alpha=A_aA^a$ in Figure \ref{fig:alpha-numerics-1}.
%

\bibliographystyle{kp}
\bibliography{Lifshitz-IR}

\end{document}